\documentstyle[prl,preprint,aps]{revtex}
\begin{document}
\draft
\title{Self-organized critical and synchronized
states in a nonequilibrium percolation model}
\author{Siegfried Clar$^1$, Barbara Drossel$^2$, and Franz Schwabl$^1$}
\address{${}^1$Institut f\"ur Theoretische Physik, \\ Physik-Department der
Technischen Universit\"at M\"unchen, \\ James-Franck-Str., D-85747 Garching,
Germany}
\address{${}^2$ Massachusetts Institute of Technology, \\ Physics Department
12-104, \\ Cambridge, MA 02139, USA}
\date{\today}
\maketitle
\begin{abstract}
We introduce a nonequilibrium percolation model which shows a self-organized
critical (SOC) state and several periodic states. In the SOC state, the
correlation length diverges slower than
the system size, and the corresponding exponent depends non universally
on the parameter of
the model. The periodic states contain an infinite cluster covering only part
of the system.
\end{abstract}


\narrowtext

During the past years, systems which exhibit self-organized criticality (SOC)
have attracted much attention, since they might explain part of the abundance
of $1/f$-noise and fractal structures in nature \cite{bak1}. Their common
features are slow driving or energy input (e.g. dropping of sand grains,
increase of strain, growing of trees) and rare dissipation events which are
instantaneous on the time scale of driving (e.g. avalanches, earthquakes,
fires).  In the stationary state, the size distribution of dissipation events
obeys a power law, irrespective of initial conditions and without the need to
fine-tune parameters (see \cite{bak3} for examples).
In general, essentially  three different types of avalanche size distributions
are possible in systems with slow energy input and instantaneous relaxation
events:
Either relaxation events occur quite frequently but release only little energy
(localized avalanches), or they occur very seldom and release a finite portion
of the systems energy (avalanches which cover the whole system), or the system
is at the borderline between both regimes and exhibits a power-law size
distribution of avalanches. In SOC systems, some mechanism pushes the system
naturally to the critical point, e.g. the local conservation of sand in the
sandpile model, or separation of time scales in the forest-fire model. A model
which has the two noncritical regimes, separated by a critical point, has been
introduced e.g. in \cite{cha1}.

In this paper, we present a model that belongs to the mentioned class of
systems, but has surprising features not known so far:
The region of small avalanches and the region of infinite
avalanches are separated by a finite region of critical behavior, where the
correlation length diverges slower than the system size.
Since criticality is not restricted to a single point, this model can be
counted among the SOC systems. In the region of infinite avalanches, the
system shows synchronization with a period that depends on the value of the
control parameter.

Our model can be obtained from the SOC forest-fire model \cite{dro1}
by taking the tree
density $\rho$ as the control parameter instead of the ratio between lightning
probability and tree growth probability $f/p$. It is defined as follows:
Initially, the fraction $\rho$ of all sites in a square lattice with $L^2$
sites are randomly chosen to be occupied. We call an occupied site ``tree''.
The remaining sites are empty. Then we repeat the following rules: (i)
Lightning strikes an arbitrary site in the system. If the site is occupied, the
whole cluster of $s$ trees which is connected to this site (by
nearest-neighbor coupling) burns down, i.e. the
trees of that cluster turn to empty sites. (ii) We grow $s$ new trees at
randomly chosen empty sites.

With these rules, the tree density $\rho$ is a globally conserved quantity. The
model therefore does not apply to any kind of burning or reaction
process. It would be more realistic to speak of ``explosions'' instead of
fires: At a randomly chosen site, an explosion takes place, and the whole
cluster connected to the explosion site is blown into the air and settles down
somewhere else in the system. Or one might think of colonies of animals which
are dispersed into all directions by some enemy or other event.
Or, from a purely
mathematical point of view, one has a nonequilibrium
percolation problem. But since the model originally was derived from a
forest-fire model, we continue to describe it in terms of trees
and fires.

In the following, we discuss the properties of the stationary state as
function of the tree density. Let $\bar s$ be the average
number of trees destroyed by a lightning stroke. For small tree densities,
there exist only small forest clusters, and consequently only few trees are
destroyed by a lightning stroke, i.e. $\bar s$ is small. With increasing tree
density, the size of the largest forest cluster increases, and $\bar s$
increases, too. There exists a critical density $\rho_c^1 \approx 0.41$, where
the size of the largest cluster and $\bar s$ both diverge. The critical
behavior close to $\rho_c^1$ can be described by exponents which are defined as
in percolation theory \cite{sta1}. The size distribution of forest clusters is
\begin{equation}
n(s) \propto s^{-\tau} {\cal C}(s / s_{\text{max}}) \, \, \, ,
\, \, \, s_{\text{max}} \propto (\rho_c^1 - \rho)^{-1 / \sigma}
\label{eva}
\end{equation}
with a cutoff function ${\cal C}$.
The fractal dimension of the clusters is defined via
\begin{equation}
R(s) \propto s^{1/\mu},
\label{ramona}
\end{equation}
where $R$ is the radius of gyration of a cluster.
The correlation length is given by $\xi \propto (\rho_c^1 - \rho)^{-\nu}$.
The quantity $\bar s$ can be considered as susceptibility and diverges as
$\bar s \propto (\rho_c^1 - \rho)^{-\gamma}$. The exponents are related via
the scaling relations $1/\sigma = \gamma / (3 - \tau) = \mu\nu$.
In our simulations, we found $\tau = 2.15(3)$, $\mu = 1.96(2)$,
$\nu = 1.20(5)$, and $\gamma = 2.09(5)$.

These exponents are identical to those of the SOC forest-fire model for $f/p
= (1 - \rho)/ \bar s \rho$, if they are appropriately redefined (see
 \cite{hen1,gra1,chr2,cla1}).
In a system which is much larger than the correlation length, the difference
between a globally conserved tree density and a tree density which is only
conserved on an average cannot be seen on length scales comparable to the
correlation length.

In the region $\rho>\rho_c^1$, which was inaccessible in the SOC forest--fire
model, one might naively expect an infinite forest cluster which spans the
whole system, as in percolation theory.
However, the values of the critical exponents for $\rho \lesssim \rho_c^1$
show that this cannot be the case here. The hyperscaling relation
$d = \mu (\tau - 1)$ is violated \cite{hen1,cla1}, which means that not every
part of the system contains a spanning cluster \cite{hen1}. This can
be seen in Fig.~\ref{states}(a), where in addition to large forest
clusters also
large clusters of empty sites exist. In contrast to ordinary percolation,
there is no homogeneously distributed set of large clusters that could
join at $\rho = \rho_c^1$ to form the infinite cluster.

Nevertheless, the largest forest cluster has to be infinitely large for
$\rho>\rho_c^1$. For $\rho_c^1<\rho<\rho_c^2 \approx 0.435$, we observe the
following behavior: The size of the largest cluster $s_{\text{max}}$
diverges, but slower than $L^2$, and the correlation length diverges
slower than $L$. We find
$s_{\text{max}} \propto L^{\phi_1}$,
$\xi \propto L^{\phi_2}$, and
$\bar{s} \propto L^{\phi_3}$ with $\rho$-dependent exponents $\phi_{1,2,3}$,
while $\tau$ and $\mu$ remain unchanged. Tab.~\ref{table1} shows the
values of the exponents for different densities.
Fig.~\ref{states}(b) shows a snapshot of the system for $\rho = 0.43$, and
Fig.~\ref{tau} shows
the size distribution of fires $s n(s)$ for different system sizes
at fixed $\rho$ (a) before and (b) after rescaling.
With Eqs.~(\ref{eva}),(\ref{ramona}) one can derive the scaling
relations $\phi_1 = \mu\phi_2$ and $\phi_3 = (3 - \tau) \phi_1$, which
are confirmed by the simulations.
For $\rho \to \rho_c^2$,
$\phi_{1,2,3}$ approach the values $\mu, 1,$ and $(3 - \tau) \mu$.

At $\rho = \rho_c^2$, the correlation length becomes
proportional to the system size, and the critical state
becomes unstable. We observe a discontinuous phase
transition to a state where the largest cluster contains
a finite portion of all trees in the system.
Fig.~\ref{states}(c) shows a snapshot of the system
for $\rho = 0.45$.
The system has five homogeneous and equally large
subphases with different
densities. The subphase with highest density contains an
infinite cluster. When lightning strikes one of the small
clusters in this state, only few trees burn down, and the state of the
system essentially remains unchanged. When lightning strikes
the infinite cluster, a large portion of all trees in
the subphase with the highest density burn down
and are regrown at empty sites all over the system.
The subphase which used to have the highest density now
has the lowest density, while the density of the other
subphases has increased. The values of the five densities
are the same as before, except that they are now associated
with different subphases. If we measure time in units of
large fires, the state of the system is periodic with
a period 5. Increasing $\rho$ further, one finds four,
three (Fig.~\ref{states}(d)), two (Fig.~\ref{states}(e)),
and finally one (sub)phase (Fig.~\ref{states}(f)), where the
infinite cluster spans the whole system.
The shape of the subphases depends on the boundary conditions.

To understand the occurrence of subphases with different densities,
we consider first a system with a density far above the
percolation threshold 0.59 for random site percolation
\cite{sta1}, and we start with a random initial state.
In the first iteration step, lightning strikes either the
infinite cluster, which consists of nearly all trees in
the system, or it strikes  one of the finite
clusters, which  are very small. In the first case, only
a few small clusters survive the fire, and most of the trees
are redistributed randomly in the system. In the second case,
only a small number of trees are redistributed.
In both cases, however, the state of the system changes little
and the new state is close to  a completely
random state. This remains true even after many iterations,
since the clusters which survived the first large fire
will burn down during one of the following large fires,
so that correlations cannot increase with time.

If we decrease $\rho$, the clusters which survive the
first large fire become larger. When the burnt trees
are regrown, part of them grow in or near these surviving
clusters, where the density then will be larger than
the mean density. In the space between these clusters
the density consequently is lower than the mean density.
If  $\rho$ is below a threshold $\rho^*$, this density
between the surviving clusters becomes smaller than
the percolation threshold 0.59. Then there exists no
infinite cluster in the system after the first time
step, and the state with one phase becomes unstable.
Evidently $\rho^* > 0.59$. In our simulations,
we find $\rho^* \simeq 0.625$.

For $\rho<\rho^*$, a stationary state with two subphases
occurs, the density of one subphase being above, the
density of the other subphase being far below the
threshold $\rho^*$. Decreasing $\rho$ further,
eventually  the density of the high-density subphase
drops below $\rho^*$, and a state with three subphases
is formed, and so on.

Let $\rho_1, \ldots, \rho_n$ be the densities in a state with n subphases,
starting with the highest density. Additionally we define the density
$\rho_{n+1}$ of the subphase which contained the infinite cluster, immediately
after the infinite cluster has been removed from the system, and before the
removed trees are regrown.
As consequence of a large fire, the different subphases just
exchange their densities, i.e.
\[
\rho_{i-1} = \rho_i + (\rho_1 - \rho_{n + 1}) \cdot
(1 - \rho_i)/(n\,(1 - \rho) + \rho_1 - \rho_{n + 1})
\]
for $i = 2, \ldots, n + 1$.
The last factor on the r.h.s. represents the fraction of trees of the
infinite cluster that are regrown in the subphase with density $\rho_i$.
We finally obtain
\begin{equation}
\frac{1 - \rho_1}{1 - \rho_2} = \frac{1 - \rho_2}{1 - \rho_3} = \ldots =
\frac{1 - \rho_n}{1 - \rho_{n+1}}. \label{siegfried}
\end{equation}
Together with $\rho = (1/n) \sum_{i=1}^n \rho_i$
we have $n$ equations for $n+1$ densities.

In our simulations, we observed a maximum number of $n = 5$ subphases. In the
following, we argue that there exists an upper limit for the number of
subphases even in an infinitely large system:  A state with $n$ subphases
is only stable if the subphase with the
second highest density has no infinite cluster.
In the limit  $n \to \infty$, we would have $\rho_1 - \rho_2 \to 0$ and
$\rho_1 \to \rho^*$. In such a state, the density of a given subphase would
increase continuously in time, until it reaches $\rho^*$. Then the infinite
cluster in this subphase would be destroyed, and the whole
cycle would restart.
We simulated
this dynamics, and measured the percentage of trees in the largest cluster as
function of the density (which itself is a function of time). This percentage
became finite at $\rho \simeq 0.595 < \rho^*$, indicating that subphases with
densities above 0.595 contain an infinite cluster. Consequently states with
$\rho_2 > 0.595$ cannot exist, which gives with Eq.~(\ref{siegfried}) a
maximum possible number of $n = 11(\pm 2)$ subphases, and a corresponding
minimum mean density $\rho = 42(\pm 0.3)\%$.

Since the transition between states with different number of subphases is
discontinuous, there are hysteresis effects. When the
density $\rho$ is decreased, a state becomes unstable when $\rho_1$ falls
below $\rho^* \simeq 0.625$, and its number of subphases increases by one.
When $\rho$ is increased, however, the subphase with the highest density
never becomes
unstable, and the transition from an $(n+1)$-subphase state to an
$n$-subphase state
takes place when the subphase with second highest density $\rho_2$ starts to
contain an
infinite cluster, i.e. when $\rho_2$ approaches $\simeq 0.595$.

To conclude, we have described a nonequilibrium percolation model which shows
several new phenomena which are unknown in equilibrium percolation. Besides
the earthquake model \cite{chr1}, this is the first model with a SOC state
where the correlation length diverges slower than the system size and the
corresponding exponent depends continuously on the parameter. In contrast
to the earthquake model, the exponent which characterizes the size
distribution of avalanches remains constant. In addition to the SOC state,
the system shows synchronization leading to periodic states with 1 to 5
subphases.

This work was supported by the Deutsche
Forschungsgemeinschaft (DFG) under Contracts No Dr 300/1-1 and
No Schw 348/7-1.

\begin{figure}
\caption{Stationary states for different densities and boundary conditions
         (trees are black, empty sites are white):
        (a) $\rho = 0.41$, $L = 4096$, periodic b.c.
        (b) $\rho = 0.43$, $L = 4096$, p.b.c.
        (c) $\rho = 0.45$, $L = 4096$, p.b.c.
        (d) $\rho = 0.50$, $L = 2048$, absorbing b.c.
        (e) $\rho = 0.55$, $L = 1024$, a.b.c.
        (f) $\rho = 0.63$, $L = 1024$, a.b.c.}
\label{states}
\end{figure}

\begin{figure}
\caption{Normalized size distribution of fires for
$\rho = 0.43$ and $L$ = 512, 1024, 2048, 4096, (a) before and
(b) after rescaling.}
\label{tau}
\end{figure}

\begin{table}
\begin{tabular}{llllll}
$\rho$   & 0.41    & 0.42    & 0.43    & 0.435 \\
\tableline
$\phi_1$ & 1.46(8) & 1.72(6) & 1.86(8) & 1.92(8) \\
$\phi_2$ & 0.79(2) & 0.92(3) & 0.97(3) & 0.99(2) \\
$\phi_3$ & 1.23(3) & 1.50(3) & 1.62(3) & 1.69(5) \\
\end{tabular}
\caption{The exponents $\phi_{1,2,3}$ of $s_{\text{max}}$, $\xi$, and $\bar{s}$
         for various densities $\rho_c^1<\rho<\rho_c^2$
         ($L = 512$, 1024, 2048, 4096)}
\label{table1}
\end{table}
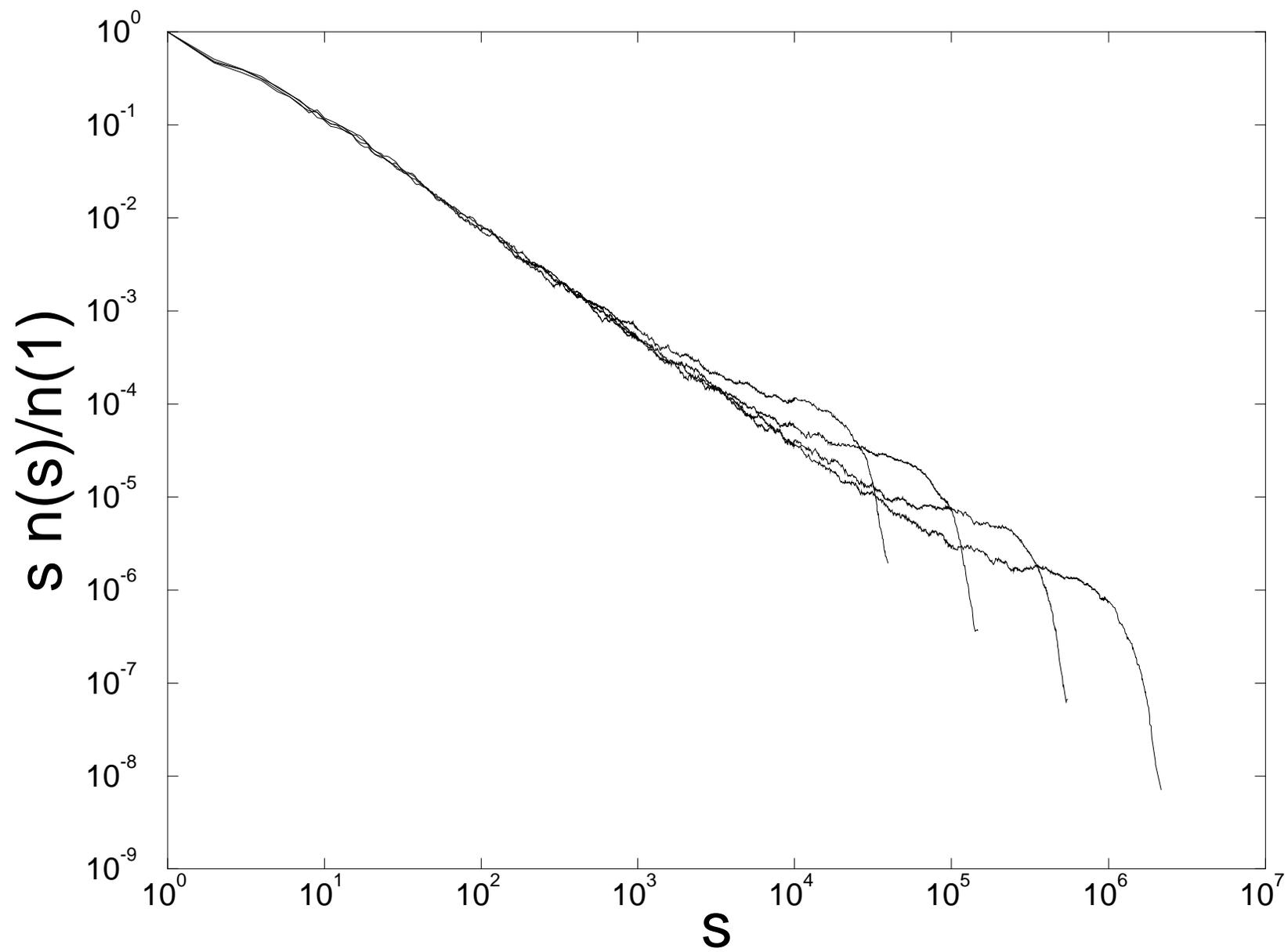
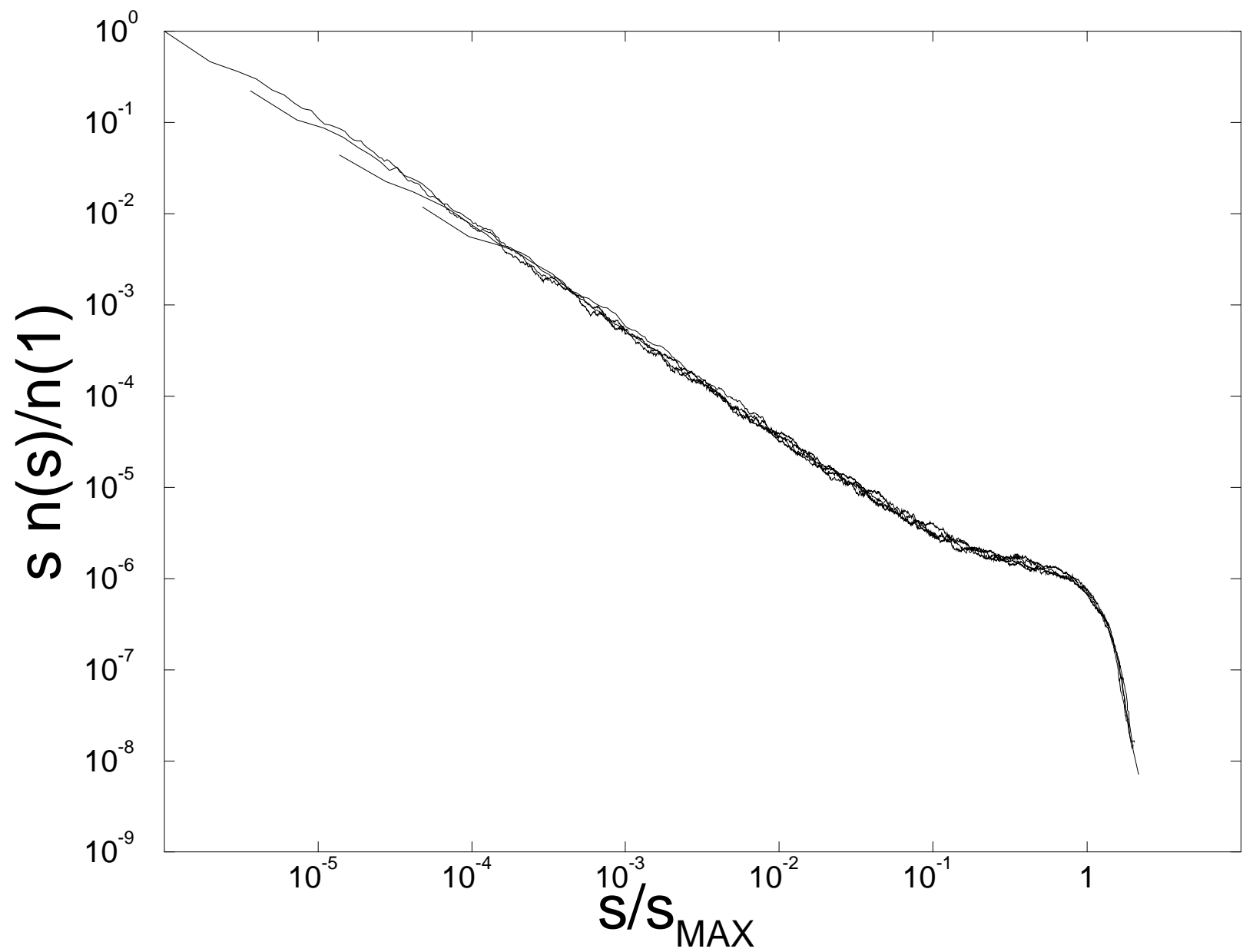
\end{document}